\documentclass[11pt]{article}
\usepackage{amsmath}
\usepackage{amssymb}
\DeclareFontFamily{U}{msb}{}
\DeclareFontShape{U}{msb}{m}{n}{ <5> <6> <7> <8> <9> gen * msbm
        <10> <10.95> <12> <14.4> <17.28> <20.74> <24.88> msbm10}{}
\DeclareSymbolFont{AMSb}{U}{msb}{m}{n}
\DeclareMathSymbol{\realset}{\mathalpha}{AMSb}{"52}
\newcommand{\mbi}{\ensuremath{\mathbf{i}}}
\newcommand{\mbj}{\ensuremath{\mathbf{j}}}
\begin{document}

\begin{flushleft}
\Large 
{\bf A central partition of molecular conformational space. \\
     I. Basic structures.}
\end{flushleft}

\vskip 0.5cm
\begin{flushleft}
\Large 
\vskip 0.5cm
Jacques Gabarro-Arpa\footnote{
Electronic Address: jga@lbpa.ens-cachan.fr \\
}
\end{flushleft}

\vskip 0.5cm
\begin{flushleft}
\large
\vskip 0.5cm
\ \ \ LBPA, C.N.R.S. UMR 8532, Ecole Normale Sup\'erieure de Cachan \\
\ \ \ 61, Avenue du Pr\'esident Wilson, 94235 Cachan cedex, France  \\
\vskip 0.5cm
\ \ \ \ \ \ \ \ \ \ \ \ \ \ \ \ \ \ \ \ and \\
\vskip 0.5cm
\ \ \ Ecole Normale Sup\'erieure, C.N.R.S. FRE 2411             \\
\ \ \ Laboratoire Interdisciplinaire de G\'eometrie Appliqu\'ee \\
\ \ \ 45, rue d'Ulm, 75230 Paris cedex, France                  \\
\end{flushleft}

\vskip  1cm
{\Large {\bf Abstract}}
\vskip 3mm

On the basis of empirical evidence from molecular dynamics simulations, 
molecular conformational space can be described by means of a partition 
of central conical regions characterized by the dominance relations between 
cartesian coordinates. This work presents a geometric and combinatorial 
description of this structure.

\newpage

{\bf Introduction}
\vskip 3mm

In previous works (Gabarro-Arpa and Revilla, 2000, 
Laboulais {\it et al.}, 2002) it was put forward the idea that the 
three-dimensional structure of proteins could be encoded into binary 
sequences. 
For a molecule with $N$ atoms in a given conformation the procedure employed 
consisted in
\begin{itemize}
\item[-] defining a procedure for enumerating the atoms, which gives
         an order relation,
\item[-] forming the set of all ordered sets of four atoms (4-tuples), 
         with size $ P = \binom{N}{4} $,
\item[-] as in the mesoscopic models of macromolecules atoms are represented 
         by pointlike structures, a 4-tuple determines 
         a 3-simplex\footnote{
          a three-dimensional (3D) polytope with four vertices.},
         since the atoms in the 4-tuples are ordered 
         a given 3-simplex can be left or right handed. Depending on 
         the simplex handedness each 4-tuple is given a sign $+/-$,
\item[-] the set of 4-tuples can also be ordered to become a sequence,  
         from it and the signs associated to each 4-tuple a sign vector 
         $\{+,-\}$$^P $ can be constructed: 
         the {\bf chirotope}\footnote{
          in this work bold faced words refer to topics that are more
          fully developed in (Rosen, 2000) and references therein.}, 
         which is the desired binary sequence.
\end{itemize}
\par
The chirotope defines an equivalence relation between conformations: two
conformations belong to the same equivalence class if they have the same
chirotope. This generates a geometrical structure in conformational space:
a partition $\mathcal{X}$ into a set of regions (cells) whose points 
(3D-conformations) have all the same chirotope.
\par
The connected components of such equivalence classes are locally compatible 
with a central conical geometry: multiplying the  $3 \times N$
cartesian coordinates of a given conformation by an arbitrary positive factor 
does not change the chirotope, since under this transformation the handedness 
of a scaled 3-simplex remains unchanged. 
Thus, in conformational space the set of points lying on a half-line starting 
at the origin all belong to the same equivalence class.
The term {\bf central} means that the vertices of the cones are at the origin.
In the following, if we talk about a {\bf partition} without further
qualifications, we mean a central partition.
\par
This simple result suggests that conformational space can be partitioned 
into a discrete set of conical cells, the structure of this partition is
encoded by the {\bf graph of regions}  
$\mathcal{T}(\mathcal{X})$, which has as vertices the set of cells of 
$\mathcal{X}$ and as edges the pairs of cells that are adjacent.
\par
Since the graph is connected, there is a graphical distance between cells
as the length of the shortest path between the two representative vertices 
in the graph.
The same distance between two equivalence classes can be defined as
a Hamming distance: the number of different signs between the chirotopes 
of the two conformations. The latter definition was first employed in
(Gabarro-Arpa and Revilla, 2000), were no geometrical interpretation
in terms of space partition was attempted.
\par
In the two works cited above the Hamming distance was used to analyze 
clusters of conformations in molecular dynamics trajectories with measurably 
good results, in these studies when compared with the classical r.m.s. 
deviation measure (Kabsch, 1978) it was seen to perform better and to be more 
robust (Laboulais {\it et al.}, 2002). This good performance can be 
qualitatively explained if the mesh that results from projecting the graph 
of regions onto the hypersurface where the system evolves, is sufficiently 
fine grained to give an accurate measure, at least in the range explored 
by molecular dynamics simulations.
\par
Thus it seems not unreasonble to give a description of conformational space
based on a central partition of conical cells. 
However, working out the set of cells derived from the chirotope turns out 
not to be pratical, so in this paper we present a partition derived from
a central hyperplane arrangement, where a set of non-coplanar hyperplanes
passing through the origin divide the space in a number 
of conical regions\footnote{
In what follows the term {\bf cone} means a region of space determined by 
a set of vectors in $\realset^N$ such that for any finite subset of vectors
it also contains all their linear combinations with positive coefficients.}.

\vskip 3mm
{\bf A central hyperplane arrangement}
\vskip 3mm

In what follows $\realset^N$ is a real affine space of $N$ dimensions, and
$\{\mathbf{e}_\mathbf{i} \}$, with $ \ 1 < \mathbf{i} < N $, are its unit 
vectors. We define the following set of vectors 
\vskip 2mm
$\mathbf{N}=\{\mathbf{n}_{\mbi\mbj}= \mathbf{e}_{\mbi}-\mathbf{e}_{\mbj} \ ,
 \ 1 \leq \mbi < \mbj \leq N \} \ \ \ \ \ \ $(1)
\vskip 2mm
\noindent
Notice that if $\mathbf{u}=(1,...,1)$ and 
$\mathbf{n}_{\mbi\mbj} \ \epsilon \ \mathbf{N}$ then 
$\mathbf{u} .\mathbf{n}_{\mbi\mbj} =0$.

\noindent
Associated with this set there is a set of central non-coplanar hyperplanes
\vskip 2mm
$\mathcal{H}_{\mbi\mbj}(p)=\{p \ \epsilon \ \realset^N : \mathbf{n}_{\mbi\mbj}.p=0\} \ \ \ \ \ \ \ \ \ \ \ \ \ \ \ \ $(2)
\vskip 2mm
\noindent
each hyperplane divides $\realset^N$ into the positive and negative hemispaces
\vskip 2mm
$\mathcal{H}_{\mbi\mbj}^+(p)=\{p \ \epsilon \ \realset^N : 
 \mathbf{n}_{\mbi\mbj}.p>0\}
 \ , \
 \mathcal{H}_{\mbi\mbj}^-(p)=\{p \ \epsilon \ \realset^N : 
 \mathbf{n}_{\mbi\mbj}.p<0\}$
\vskip 2mm
\noindent
so the hyperplane arrangement determines a partition $\mathcal{P}$ 
of $\realset^N$ into a set a set of convex regions (cells), 
where each cell $\mathcal{C} \ \epsilon \ \mathcal{P}$ 
can be characterized by an antisymmetric $N \times N$ sign matrix 
$\mathcal{V}$, such that if $p \ \epsilon \ \mathcal{C}$ then
\vskip 2mm

$\mathcal{V}_{ii} = 0 \ \ , \ \
 \mathcal{V}_{ij} = \left\{\begin{array}{rr}
 + & \mbox{ $p \ \epsilon \ \mathcal{H}_{\mbi\mbj}^+$ } \\
 0 & \mbox{ $p \ \epsilon \ \mathcal{H}_{\mbi\mbj}  $ } \\
 - & \mbox{ $p \ \epsilon \ \mathcal{H}_{\mbi\mbj}^-$ } \\
\end{array}\right. $ , \ \
$\mathcal{V}_{ji} =  \left\{\begin{array}{rr}
 - &  \\
 0 &  \\
 + &  \\
\end{array}\right. $
$ \ \ \forall \ i < j \ \ \ \ \ \ \ $(3)

\vskip 2mm
The geometrical meaning of the sign matrix can be easily  deduced
from the following example: let $p^{a} , p^{b}  \ \epsilon \ \realset^N $ 
be two points with coordinates
\vskip 2mm
$p^{a} = (1,2,3,4,5,...,N) $ \\
\indent
$p^{b} = (1,2,4,3,5,...,N) $
\vskip 2mm
\noindent
obviously $\mathcal{V}^{a} = \{ \mathcal{V}_{ij}^{a} = + \ ,
\ \forall \ i < j \}$ is the sign matrix of $p^{a}$. Now $p^{b}$
has the same matrix except that $\mathcal{V}_{34}^{b} = - \ $.
Thus, for any point $p$ the sign matrix encodes the pairwise dominance 
relations between its coordinates
\vskip 2mm

$\mathcal{V}_{ij} =  \left\{\begin{array}{rr}
 + & \mbox{ $p_{i} < p_{j} $ } \\
 0 & \mbox{ $p_{i} = p_{j} $ } \\
 - & \mbox{ $p_{i} > p_{j} $ } \\
\end{array}\right. $
$ \ \ \forall \ i < j $

\vskip 2mm
\noindent
In the above example notice also that 
$\mathbf{n}_{\mathbf{3}\mathbf{4}} = p^{b} - p^{a} \ \ \ (4)$.
\par
Let $\pi(N)$ be the set of points in $\realset^N$ whose coordinates are 
the permutations of the sequence $\{1,2,3,4,5,...,N\}$, no two points 
in this set have the same $\mathcal{V}$ matrix, and since it encodes 
the complete set of dominance relations between coordinates, there is a one 
to one correspondence between $\pi(N)$ and $\mathcal{P}$, making a total of 
$N!$ cells in $\mathcal{P}$.

\vskip 3mm
{\bf The graph of regions of the arrangement}
\vskip 3mm

In order to study the graph of regions $\mathcal{T}(\mathcal{P})$,
it is important to notice that $\mathcal{V}$ is the incidence matrix of 
an {\bf acyclic tournament} (Moon, 1968).
\par
Tournaments are directed graphs such that between any two nodes 
there is always an arc (see example in fig. 1), if $v_i$ and $v_j$ are two 
nodes, $\mathcal{V}_{ij} = +$ if the arc goes from $i$ to $j$,
we say that $v_i$ {\bf dominates} $v_j$; otherwise $\mathcal{V}_{ij} = -$
and $v_i$ is dominated by $v_j$.
\par
The {\bf acyclic}\footnote{
All along this work tournaments are implicitly assumed to be acyclic.}
 qualifier is because there are no directed cycles in the
graph (as can be seen in fig. 1). This is a particularity of the tournaments 
that characterize the cells of $\mathcal{P}$: for any permutation 
there are always two nodes called the {\bf source} and the {\bf sink} 
respectively, the former dominates all other nodes and the latter 
is dominated by every node in the graph (nodes $3$ and $2$ in fig. 1, 
respectively). Moreover it is a {\bf centrally symmetric hierarchical} 
structure: 
\begin{itemize}
\item the graph that results from reversing all the arcs is also acyclic, 
\item deleting a node always gives a subtournament.
\end{itemize}
\par
This tells us that each $N$-dimensional cell in $\mathcal{P}$ has exactly 
$N-1$ neighbours, since there are exactly $N-1$ arcs in a tournament
that can be reversed without creating a directed cycle.
These are the arcs joining nodes whose score differs by 1 (see the legend 
of fig. 1).
\par
$\mathcal{T}(\mathcal{P})$ can be obtained by joining with a line segment 
the points in $\pi(N)$ that are in adjacent cells, the result is the 
1-{\bf skeleton} of a convex polytope: the $N$-{\bf permutohedron} 
or $\Pi_{N-1}$ (Schoute, 1911).
\par
The study of the faces of $\Pi_{N-1}$ is an essential part in our study 
of $\mathcal{P}$, since it allows  conformations and groups of conformations
to be accurately located within $\Pi_{N-1}$.

\vskip 3mm
{\bf The faces of $\Pi_{N-1}$ }
\vskip 3mm

Central to this construction is the duality between the faces of $\Pi_{N-1}$ 
and the cells of $\mathcal{P}$: $k$-faces and cells of dimension $N-k$ 
lie in orthogonal linear subspaces. The sign matrix of lower dimensional cells 
has zeros in the entries corresponding to hyperplanes that contain the cell, 
as defined in (3), this matrix can be represented by incomplete 
tournaments: these are digraphs where the arcs corresponding to the zero 
entries have been deleted (see fig. 2).
\par
Incomplete tournaments can be seen as patterns: we say that a given tournament 
matches a {\it pattern} if both graphs have the same {\bf order} and if the 
pattern is a subgraph of the tournament.
\par
The simplest non-trivial faces in the hierarchy are the 1-faces: edges 
that join adjacent vertices ($0$-faces). As we have seen, 
adjacent vertices differ in that they exchange the value of two coordinates,
say $\mbi$ and $\mbj$, and the edge is parallel to 
the vector $\mathbf{n}_{\mbi\mbj}$ (4), which is perpendicular to the 
hyperplane $\mathcal{H}_{\mbi\mbj}$. This hyperplane contains the 
$(N-1)$-dimensional boundary cell that separates the $N$-dimensional cells 
of the vertices, accordingly its sign matrix has 
$\mathcal{V}_{\mbi\mbj}=\mathcal{V}_{\mbj\mbi}=0$. 
\par
The pattern of fig. 2a, where the arc between $v_2$ and $v_6$ is missing,
matches exactly two tournaments that represent the vertices of the
edge segment, also the complement graphs\footnote{
Given a tournament $T$ and a pattern $P$ the {\it complement graph} 
is a graph with the edges that are in $T$ but not in $P$ and the vertices that
are in those edges (see fig. 3).} 
 of these vertices is a set of lower order tournaments 
that encode the vertices of a lower dimensional permutohedron. In our case 
we have 
two order 2 tournaments (one can be seen in fig. 3a), that represent the 
permutations of the sequence $\{56\}$ : the associated permutohedron is a 
line segment.
\par
On the other extreme lets look at the $(N-2)$-faces. Notice that for 
the hyperplane arrangement described above if we construct the vector
\vskip 2mm
$ \mathbf{u}^\alpha = \{ 
  \mathbf{u}^\alpha_\alpha = 1-N \ , \
  \mathbf{u}^\alpha_i = 1 \ , \ 
i \neq \alpha \ , \ 
1 \leq i      \leq N  \} \ \ \ \ \ \ \ $(5) \\
\noindent
where $1 \leq \alpha \leq N$; the set of vectors
\vskip 2mm
$ \mathbf{N}^\alpha = \{ 
  \mathbf{n}_{\mbi\mbj} \ \epsilon \ \mathbf{N} : 
   1 \leq \mbi < \mbj \leq N  \ , \ 
  \mbi \neq \alpha \ , \ \mbj \neq \alpha \} \ \ \ \ \ \ \ \ \ \ $(6)
\vskip 2mm
\noindent
and the subset of hyperplanes
\vskip 2mm
$ \mathcal{H}^\alpha = \{ 
  \mathcal{H}_{\mbi\mbj}(p) : \mathbf{n}_{\mbi\mbj}.p = 0 \ , \
  \mathbf{n}_{\mbi\mbj} \ \epsilon \ \mathbf{N}^\alpha\   \ , \
  p \ \epsilon \ \realset^N \} $
\vskip 2mm
\noindent
we have $ \mathbf{u}^\alpha.\mathbf{n}_{\mbi\mbj}=0 $ for all
$ \mathbf{n}_{\mbi\mbj} \ \epsilon \ \mathbf{N}^\alpha $. This means that
the vectors in $\mathbf{N}^\alpha$ are all in the $(N-1)$-hyperplane 
$\mathbf{u}^\alpha.p = 0$, consequently 
\begin{itemize}
\item the hyperplanes in $\mathcal{H}^\alpha$ all have a common intersection: 
      a 2-hyperplane parallell to $\mathbf{u}^\alpha$, 
\item the set of vertices in the $N$-cells adjacent to this 2-cell
      all lie in a $(N-2)$-hyperplane: they are the vertices of a $(N-2)$-face,
\item as can be deduced from (5) and (6), the only arcs present in the pattern 
      of this face are those that connect node $\alpha$ to the other nodes, 
\item node $\alpha$ is either a source or a sink.
\end{itemize}
\par
For $N=6$ we have the example pattern of fig. 2b, fig. 3b shows 
the complement graph which is a $N=5$ tournament, the set of all complement
graphs encodes the permutations of the set $\{12345\}$, and hence 
the corresponding face is a $\Pi_4$ polytope.
This tells us that a total of $2 N$ faces of $\Pi_{N-1}$ are $\Pi_{N-2}$ 
polytopes.
\par
Before proceeding further we must introduce the basic notion of {\bf product
polytope}. If $P$ is a polytope in $\realset^p$ and $Q$ a polytope
in $\realset^q$ then the product polytope $P \times Q$ is defined as  the set 
of all vertices $(x,y) \ \epsilon \ \realset^{p+q}$ 
such that $x \ \epsilon \ \realset^p$ is a vertex of $P$ and 
$y \ \epsilon \ \realset^q$ is a vertex of $Q$.
Examples of product polytopes are: the square which is the product of two
segments (two polytopes of dimension 1). The cube which is the 
product of a square by a segment, more generally the prisms, which are the
product of a polygon (or polytope) by a segment.
\par
The example pattern from fig. 2c encodes a product polytope: the set of 
compatible subtournaments formed by vertices $v_1$, $v_3$, $v_4$ and $v_5$, 
encode a $\Pi_3$, and are independent from the subtournament formed
by  $v_2$ and $v_6$ which encodes a segment (or $\Pi_1$). Thus there are 
$N \times (N-1)$ $(N-2)$-dimensional faces which are prisms joining two
$\Pi_{N-3}$ from adjacent $\Pi_{N-2}$ faces.
\par
It can be easily seen that all the faces from this polytope are either
permutohedrons or products of permutohedrons, for instance the polytope 
encoded by the pattern of fig. 2d is a $\Pi_1 \times \Pi_1 \times \Pi_1$, 
that is: a cube.
\par
Notice that for the product of permutohedrons the complement graphs 
(see figs. 3c and 3d) are not connected.

\vskip 3mm
{\bf The face lattice of $\Pi_{N-1}$ }
\vskip 3mm

The differences among incomplete tournaments, when we disregard the identity 
of the nodes, arise from the topology of the graph: number of edges and nodes,
and the connectivity. 
We can define an operation on patterns that consists in renunbering
the nodes so that the score never decreases upon increasing the node number. 
Renumbered patterns are stripped from the complications that arise from 
permuting equivalent nodes, a classification of these objects based on 
topological differences, is far more simple while keeping their essential 
characteristics, it results in a comprehensive synthetic view of the face 
arrangement. Introducing the permutations between equivalent nodes is an 
unnecesary complication that can always be worked out in a later stage.
\par
The set of equivalence classes obtained upon renumbering is isomorph 
to the set $\mathcal{L}$ of partitions of the sequence $\{1,2,3,...,N\}$ 
into subsets of consecutive integers. 
The correspondence is established as follows
\begin{itemize}
\item for each incomplete tournament form a sequence with 
      the partial scores arranged in ascending order
\item divide this sequence into subsets of identical partial scores
\item replace each element in a subsequence of identical scores by the 
      the corresponding node number in the renumbered sequence.
\end{itemize}
For instance from the graph of fig. 2b we form the sequence $111116$,
whose corresponding partition is $(12345)6$. Likewise, for figs. 2a, 2c 
and 2d the sequences $123455$, $111155$ and $113355$,
give the partitions $1234(56)$, $(1234)(56)$ and $(12)(34)(56)$ respectively.
These partitions represent classes of polytopes that are combinatorially
equivalent, there are $N!/(n_1! \times n_2! \times ... )$ elements in each 
class, $n_i$ being the number of elements in each subset.
Notice that $(56)$, for instance, represents the set of permutations 
of the sequence $\{56\}$
\par
There is a partial order in the set of partitions, it is based on
containment: we say that a partition set $x$ is contained in $y$ 
($x \subset y$), if each subset in $x$ is either identical to a subset 
of $y$ or it is a subset of some subset in $y$.
Thus for the above example: $1234(56) \subset (12)(34)(56) \subset 
(1234)(56)$, also $(1234)56 \subset (1234)(56)$ and $(1234)56 \subset 
(12345)6$, but $(1234)(56) \not\subset (12345)6$.
\par
It can be shown that the partially ordered set ({\bf poset}) $\mathcal{L}$ 
thus defined is a {\bf lattice}, that is: for all pairs 
$x \ , \ y \ \epsilon \ \mathcal{L}$ there is a least upper bound
and a greatest lower bound.
\par
The lattice poset $\mathcal{L}$ for $N=6$ is represented in fig. 4, 
each element represents a class of faces of $\Pi_5$, they are arranged 
in five rows, the faces in a row have the same dimension which increases 
from 0 in the bottom row to 5 at the top. There are 
$ \binom{N-1}{d} \ (0 \leq d \leq N)$, elements in each row.
\par
It should be noticed from fig. 4 the hierarchical structure of $\mathcal{L}$:
each interval\footnote{
An {\bf interval} is a subposet which contains all elements $z$ such that
$x \subseteq z \subseteq y$.}
between a given element in the lattice and the minimal element $123456$
is also a lattice. Which is an expected result: the face lattice of any face
is in the face lattice.

\vskip 3mm
{\bf A partition of conformational space }
\vskip 3mm

The partition discussed in the previous sections is based on the dominance
relations among the coordinates of points in an $N$-dimensional space, 
in conformational space the coordinates of each point are the coordinates 
of a set of $N$ points in 3D cartesian space, as 3D cartesian 
coordinates are independent of each other it would make little sense to 
translate automatically the partition described above to a $3 \times 
N$-dimensional space, instead we propose the partition $\mathcal{P}^3$ which 
is the union of three separate partitions: $\mathcal{P}_x$, $\mathcal{P}_y$ 
and $\mathcal{P}_z$, that encode the dominance relations among the $x$, $y$ 
and $z$ coordinates of the set of points respectively.
$\mathcal{P}_x$, for instance, is generated 
by the set of hyperplanes 
\vskip 2mm
$\mathcal{H}^x_{\mbi\mbj}(p)=\{p \ \epsilon \ \realset^{3 \times N} : 
 \mathbf{n}^x_{\mbi\mbj}.p=0\}$
\vskip 2mm
\noindent
with a set of normal vectors defined as 
\vskip 2mm
$\mathbf{N}^x=\{\mathbf{n}^x_{\mbi\mbj}= 
 \mathbf{e}^x_{\mbi}-\mathbf{e}^x_{\mbj} \ , \ 
 1 \leq \mbi < \mbj \leq N \}$
\vskip 2mm
\noindent
where the $\mathbf{e}^x$ are the unit vectors in $\realset^{3 \times N}$ 
of the $x$ coordinates of the 3D points. 
\par
$\realset^{3 \times N}$ can be seen as a product space 
$\realset^N \times \realset^N \times \realset^N$, with each factor
harboring the $x$, $y$ and $z$ coordinates of the set of points. Thus, as the 
dual polytope of $\mathcal{P}_x$, for instance, is $\Pi_{N-1}$, obviously 
the dual of $\mathcal{P}^3$ will be 
$\Pi^3_{N-1} = \Pi_{N-1} \times \Pi_{N-1} \times \Pi_{N-1}$, its face poset
can be worked out from the observation that $\Pi^3_{N-1}$ is a $(3N-3)$-face
of $\Pi_{3N-1}$. See for example the symmetric class of faces $(12)(34)(56)$ 
in fig. 4, the poset of $\Pi^3_1$ is the interval $123456-(12)(34)(56)$.
\par
Now the question that arises is: how well do 3D $N$ point sets 
arising from the vertices of $\Pi^3_{N-1}$ relate to the actual conformations 
of macromolecules ?
\par
An alternative representation of permutations is as {\bf $0/1$ matrices}, 
these are objects whose only entries are $0$s and $1$s with the entry $1$ 
occuring exactly once in each column. As an example to the permutation 
encoded by the tournament of fig. 1 it corresponds the $0/1$ matrix

\vskip 3mm
$ \ \ \ \ \ \ \ \ \ \ \ \ \ \ \ \ \ \ \ \ \ \ \ \ \ \ \ \ \ \ \ \ \ \  \ \ \ \
  \ \ \ \ \ \
  \begin{pmatrix} 
       0 & 1 & 0 & 0 & 0 & 0 \\
       0 & 0 & 0 & 0 & 0 & 1 \\
       0 & 0 & 0 & 0 & 1 & 0 \\
       1 & 0 & 0 & 0 & 0 & 0 \\
       0 & 0 & 0 & 1 & 0 & 0 \\
       0 & 0 & 1 & 0 & 0 & 0 
  \end{pmatrix} $
\vskip 3mm

\noindent
likewise the coordinates of a vertex in $\Pi^3_{N-1}$ can be encoded by a
three-dimensional $0/1$ matrix, which can be regarded as a cubic lattice
with only one site occupied per row in every dimension. 
Analogously, we can imagine in 3D cartesian space an $N$-point set 
embedded in a cubic lattice with cell spacing of $1$
spanning a rectangle between $1$ and $N$ in every dimension
with the points located at the intersections such that there is only one 
point in every row in any dimension.
\par
We can compare in fig. 5 the 3D stereoviews of the HIV-1 integrase 
catalytic core C$_\alpha$ skeleton (fig. 5a)\footnote 
{residues 50-212 of the integrase (Maignan {\it et al.}, 1998).}, 
with the 3D representation, fig. 5b, of the corresponding $\Pi^3_{162}$ 
vertex within the $\mathcal{P}^3$ cell. 
Altough fig. 5b appears to be somewhat deformed with respect to fig. 5a, 
all the characteristic folding patterns: $\alpha$-helices, $\beta$-sheets, 
turns ... appear to be conserved.
\par
This means that a lot of the 3D structure is encoded by the set of dominance
relations among the cartesian coordinates of individual atoms.

\vskip 3mm
{\bf Conclusion}
\vskip 3mm

Most of the time conformational space is referenced as an abstract paradigm
too complex to be understood.
The aim of this work is to show that the geometry of conformational space 
is not beyond the reach of mathematical intuition: with the help of 
adequate mathematical structures its sheer complexity can be brought 
to tractable dimensions, and it can be done with existing and well 
understood mathematical tools.
\par
The model developped here offers a number of interesting possibilities 
\begin{itemize}
\item the structural diversity of a macromolecule can be explored by means
      of combinatorial patterns
\item the classification of conformations can give a catalog of structures 
\item graphical paths can be used to determine and explore the paths between 
      any two conformations
\item its hierarchical structure makes it modular
\end{itemize}
There are shortcomings too: the present model shows a loss of precision in 
the 3D-structures obtained; but this should not be a major problem:
\begin{itemize}
\item precision can be recovered with the help of {\it ad hoc} methods.
      Optimization of structures within a cell should not be difficult
\item there is no limit to the refinements that can be introduced into
      this basic model, in particular it should not be hard to build
      smaller cells, or to cut the existing ones into finer slices.
\end{itemize}
\par
The possibilities offered by the model will be the subject of the 
forthcoming works.

\newpage
{\bf References}
\vskip 3mm

\noindent
Gabarro-Arpa, J. and Revilla, R. (2000) {\it Comput. and Chem.}
{\bf 24}, 693-698.
\vskip 2mm
\noindent
Kabsch, W. (1978) {\it Acta Cryst.} {\bf A34} 827-828.
\vskip 2mm
\noindent
Laboulais, C., Ouali, M., Le Bret, M. and Gabarro-Arpa, J. (2002)
{\it Proteins: Structure, Function, and Genetics} {\bf 47} 169-179.
\vskip 2mm
\noindent
Maignan, S., Guilloteau, J. P., Zhou-Liu, Q., Cl\'ement-Mella, C., Mikol, V. 
(1998) {\it J. Mol. Biol.} {\bf 282} 359-368.
\vskip 2mm
\noindent
Moon, J. W. (1968) {\it Topics on Tournaments }, Holt, Rinehart and Winston,
New York.
\vskip 2mm
\noindent
Rosen, K. editor in chief (2000) {\it Handbook of Discrete and
Combinatorial Mathematics} CRC Press, New York.
\vskip 2mm
\noindent
Schoute, P. H. (1911) Verhandelingen der Koninklijke Akademie 
van Wetenschappen te Amsterdam, Deel {\bf 11}, No. 3,
Johannes Muller, Amsterdam, 1-87.

\newpage
{\LARGE {\bf Legends of Figures}}

\vskip 6mm
{\bf Figure 1}
\vskip 4mm
$N=6$ tournament corresponding to the sign matrix 

\vskip 3mm
$ \mathcal{V} =  { 
      \begin{pmatrix} 
       0 & - & + & + & - & - \\
       + & 0 & + & + & + & + \\
       - & - & 0 & - & - & - \\
       - & - & + & 0 & - & - \\
       + & - & + & + & 0 & - \\
       + & - & + & + & + & 0 
      \end{pmatrix} } $
\vskip 2mm

The {\bf score} of a node is the number of nodes it dominates plus 1 (in
order to establish a correspondence with permutations). It is annotated above 
each node in the figure.

\vskip 4mm
{\bf Figure 2}
\vskip 4mm

Example incomplete tournaments for $N=6$ matching the tournament of fig. 1. 
Their respective sign matrices are
\vskip 3mm
$ \mathcal{V}^a =  { 
      \begin{pmatrix} 
       0 & - & + & + & - & - \\
       + & 0 & + & + & + & 0 \\
       - & - & 0 & - & - & - \\
       - & - & + & 0 & - & - \\
       + & - & + & + & 0 & - \\
       + & 0 & + & + & + & 0 
      \end{pmatrix} }  \quad  \quad
  \mathcal{V}^b =  { 
      \begin{pmatrix} 
       0 & - & 0 & 0 & 0 & 0 \\
       + & 0 & + & + & + & + \\
       0 & - & 0 & 0 & 0 & 0 \\
       0 & - & 0 & 0 & 0 & 0 \\
       0 & - & 0 & 0 & 0 & 0 \\
       0 & - & 0 & 0 & 0 & 0 
      \end{pmatrix} } $
\vskip 2mm
$ \mathcal{V}^c =  { 
      \begin{pmatrix} 
       0 & - & + & + & - & - \\
       + & 0 & + & + & + & 0 \\
       - & - & 0 & 0 & - & - \\
       - & - & 0 & 0 & 0 & - \\
       + & - & + & 0 & 0 & - \\
       + & 0 & + & + & + & 0 
      \end{pmatrix} }  \quad  \quad
  \mathcal{V}^d =  { 
      \begin{pmatrix} 
       0 & - & + & + & 0 & - \\
       + & 0 & + & + & + & 0 \\
       - & - & 0 & 0 & - & - \\
       - & - & 0 & 0 & - & - \\
       0 & - & + & + & 0 & - \\
       + & 0 & + & + & + & 0 
      \end{pmatrix} } $

\vskip 4mm
{\bf Figure 3}
\vskip 4mm

Complement graphs of the tournament in fig. 1 with respect to the patterns in 
fig. 2.

\vskip 4mm
{\bf Figure 4}
\vskip 4mm

Poset $\mathcal{L}$ of the partitions of the sequence $(123456)$ into 
subsets of consecutive integers.
The bold letters above some partitions refer to the incomplete tournaments 
in fig. 2.

\vskip 4mm
{\bf Figure 5}
\vskip 4mm

\begin{itemize}
\item[{\bf a})] Stereo drawing of the HIV-1 integrase catalytic core 
                C$_\alpha$ skeleton (residues 50-212 of the integrase,  
                Maignan {\it et al}., 1998).
\item[{\bf b})] Stereo drawing of the related vertex in $\Pi^3_{162}$.
\end{itemize}

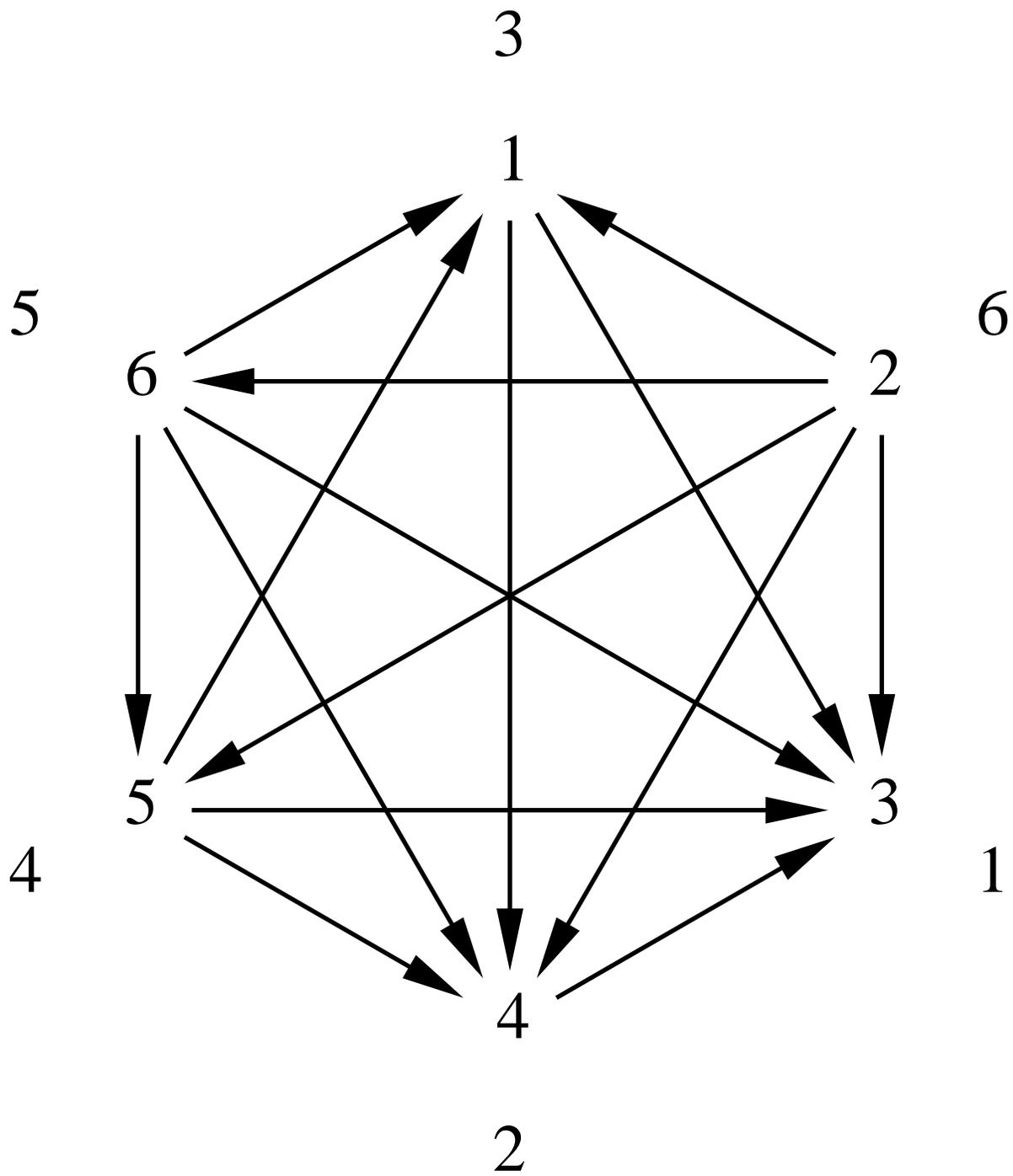
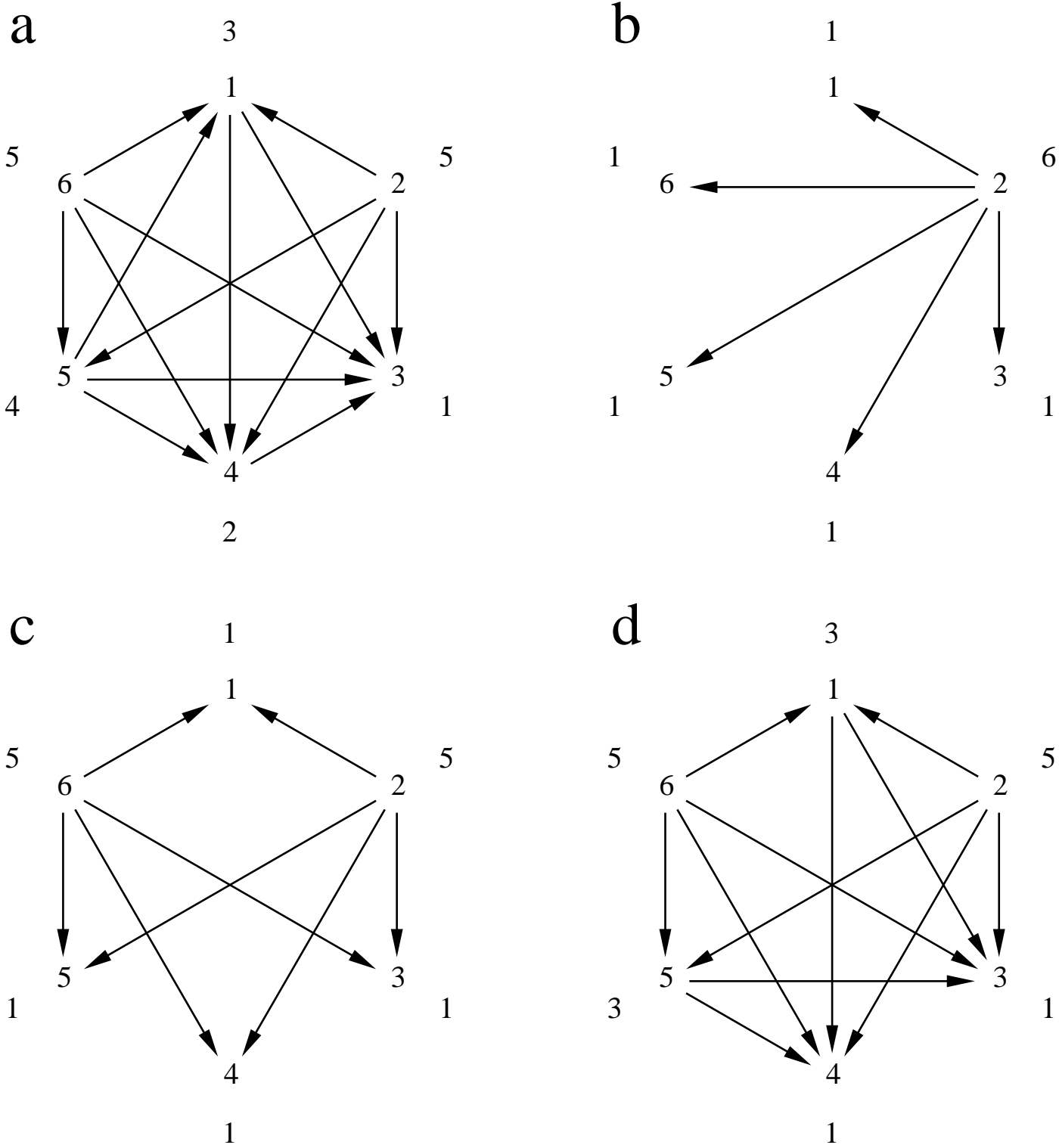
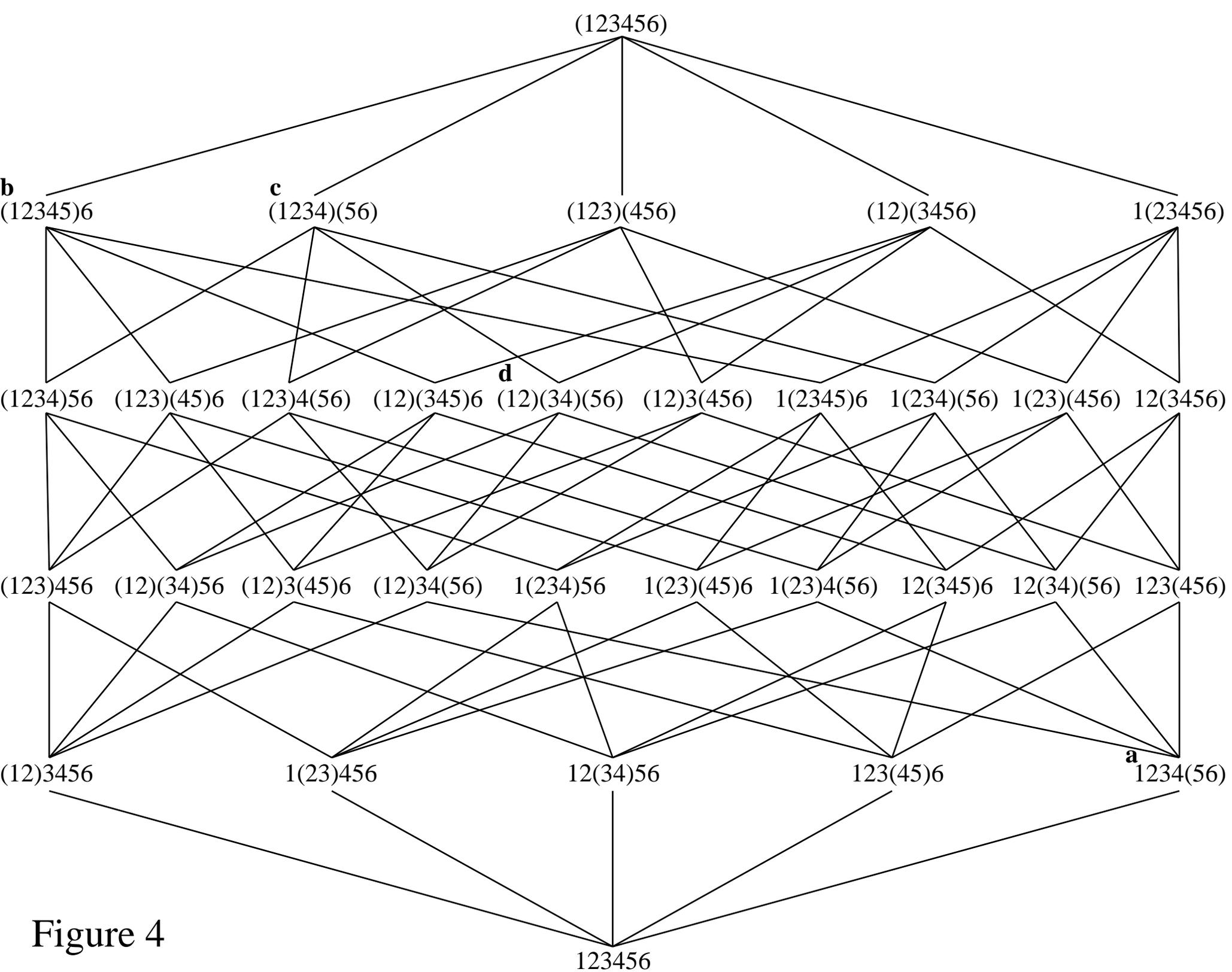
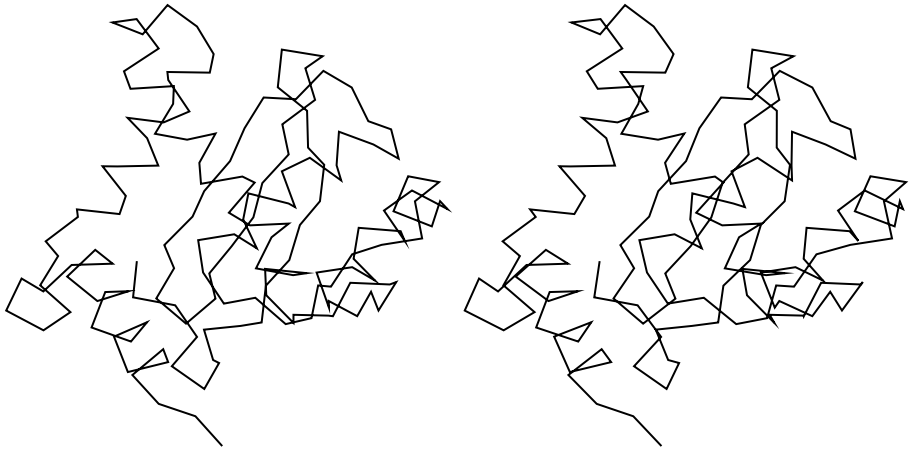
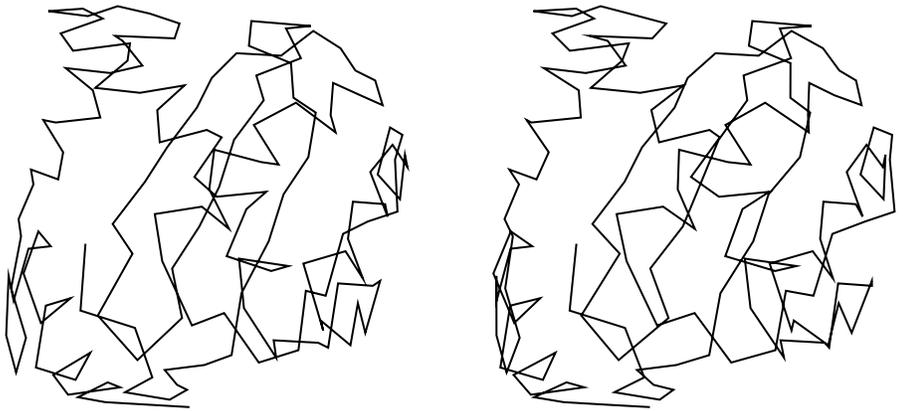
\end{document}